# 2-Dimensional Finite Impulse Response Graph-Temporal Filters

Elvin Isufi, Geert Leus and Paolo Banelli

*Abstract*—Finite impulse response (FIR) graph filters play a crucial role in the field of signal processing on graphs. However, when the graph signal is time-varying, the state of the art FIR graph filters do not capture the time variations of the input signal. In this work, we propose an extension of FIR graph filters to capture also the signal variations over time. By considering also the past values of the graph signal, the proposed FIR graph filter extends naturally to a 2-dimensional filter, capturing jointly the signal variations over the graph and time. As a particular case of interest we focus on 2-dimensional separable graph-temporal filters, which can be implemented in a distributed fashion at the price of higher communication costs. This allows us to give filter specifications and perform the design independently in the graph and temporal domain. The work is concluded by analyzing the proposed approach for stochastic graph signals, where the first and second order moments of the output signal are characterized.

## I. INTRODUCTION

Signal processing on graphs emerged recently as a tool to extend classical signal processing concepts from time and image signals to signals that reside on the vertices of an irregular graph. The breakthrough in this area is the definition of the graph Fourier transform (GFT) [1]–[3], which extends the analysis of graph signals to the graph frequency domain. By having a specific definition of graph frequency, graph filters emerged as a basic building block to process the spectral content of graph signals. Graph filters have been used in applications like data classification and customer behavior prediction [1], signal denoising and smoothing [4], [5], solving consensus problems [6], anomaly and event boundary detection [7], [8], to name a few. Finite impulse response (FIR) graph filters appeared first with the property of having a polynomial frequency response, meaning they can be easily implemented in a distributed manner in the node domain [1], [9]. Secondly, infinite impulse response (IIR) graph filters were proposed due to the necessity to achieve better interpolation or extrapolation properties around the known graph frequencies [8], [10], [11]. However, the aforementioned works focus mainly on time-invariant graph signals, whereas time can carry extra information, for example financial time series of companies or goods in a stock market, temperature measurements taken continuously in time by a sensor network, or political popularity in social media. For these cases we can be interested in computing predictions, statistics or make inferences on this signal. This can potentially be more accurate when the time dependency of the signals is taken into account. Such aspects are catching attention recently, also in the graph signal processing area [11], [12].

In [11], an autoregressive moving average (ARMA) is proposed which has the ability to process jointly the graph and temporal variations of the signal. However, in the design process stability issues arise. To avoid the latter problem related to IIR filters, we present an extension of FIR graph filters to capture also the time-variations of the graph signal. By incorporating also a temporal memory while computing the filter output, the well-known FIR graph filters extend to 2D filters operating jointly on the graph and temporal spectral domain. In contrast to [11], the proposed 2D filter has more degrees of freedom to approximate a desired frequency response. Further, as a particular case we analyze the class of separable 2D frequency responses. This property allows us to give filter specifications and perform the design independently in the graph and temporal domain.

The paper is concluded by analyzing the proposed approach for a time-varying random process over the graph. For this case, we calculate in closed form the first and second order moments of the filter output and we show that, in the mean, the proposed filter behaves as the same filter operating on a deterministic signal being the mean of the graph process.

## II. PRELIMINARIES

Let us consider an undirected and connected graph $\mathcal{G}$ of $N$ nodes. We indicate with $\boldsymbol{x} \in \mathbb{R}^N$ the graph signal and with $\boldsymbol{L}$ the graph Laplacian [3]. The GFT $\hat{\boldsymbol{x}}$ of $\boldsymbol{x}$ and its inverse are calculated as

$$\hat{x}_i = \langle \boldsymbol{x}, \boldsymbol{\phi}_i \rangle, \text{ and } x_i = \sum_{n=1}^{N} \hat{x}_n \boldsymbol{\phi}_n(i), \quad (1)$$

respectively, where $\langle \cdot \rangle$ denotes the inner product, $\boldsymbol{\Phi} = [\boldsymbol{\phi}_1, \ldots, \boldsymbol{\phi}_N]^\top$ are the Laplacian's eigenvectors and $\boldsymbol{\phi}_n(i)$ is the $i$th entry of $\boldsymbol{\phi}_n$. The corresponding eigenvalues $\{\lambda_n\}_{n=1}^{N}$ form the graph frequencies. We present our results for a Laplacian[1] matrix $\boldsymbol{L}$, where we only require $\boldsymbol{L}$ to be symmetric and local: for all $i \neq j$, $\boldsymbol{L}_{ij} = 0$ whenever the nodes $u_i$ and $u_j$ are not neighbours and $\boldsymbol{L}_{ij} = \boldsymbol{L}_{ji}$ otherwise.

A graph filter $\boldsymbol{H}$ is defined as a linear operator that acts on a graph signal $\boldsymbol{x}$ by shaping its spectrum as

$$\boldsymbol{H}\boldsymbol{x} = \sum_{n=1}^{N} H(\lambda_n) \langle \boldsymbol{x}, \boldsymbol{\phi}_n \rangle \boldsymbol{\phi}_n. \quad (2)$$

The graph frequency response $H : \{\lambda_n\}_{n=1}^{N} \to \mathbb{R}$ controls how much $\boldsymbol{H}$ amplifies the signal component of each graph frequency (suppose for now $\langle \boldsymbol{x}, \boldsymbol{\phi}_n \rangle \neq 0$)

$$H(\lambda_n) = \langle \boldsymbol{H}\boldsymbol{x}, \boldsymbol{\phi}_n \rangle / \langle \boldsymbol{x}, \boldsymbol{\phi}_n \rangle. \quad (3)$$

Given a desired graph frequency response $H^*(\lambda)$, the filter coefficients are found by solving a linear system when the graph and thus the frequencies $\lambda_n$ are known [1], [3], [13]. In case the graph is not known, we can approximate $\boldsymbol{H}$ by using a $K$-th order polynomial of $\boldsymbol{L}$ and the filter output is

$$\boldsymbol{y} = \Big(a_0 \boldsymbol{I} + \sum_{k=1}^{K} a_k \boldsymbol{L}^k\Big)\boldsymbol{x}. \tag{4}$$

---

[1]Note that the core idea can be applied also to directed graphs using the adjacency matrix instead of the Laplacian.

To such a design approach is commonly referred as universal design [9], [11]. An order-$K$ FIR graph filter (FIR$_K$) can be computed distributively, since $\boldsymbol{L}^K \boldsymbol{x} = \boldsymbol{L}(\boldsymbol{L}^{K-1}\boldsymbol{x})$ and each node can compute the $K$-th term from the values of the $(K-1)$-th term in its neighbourhood.

## III. FIR GRAPH FILTERS WITH TIME-VARYING INPUT

In this section, we present the recursions that implement 2D graph-temporal filters. We start building our approach from an intuitive extension of (4), which can also be implemented distributively with the same computational efforts. Then, we move to the more general approach, which requires $K$ times more data exchanges and computational power to implement a 2D FIR filter, but it offers the more complete 2D transfer function. Further, we present two particular subclasses, which are able to implement casual 2D FIR filters and separable filters in the graph and temporal domain, respectively.

**2D FIR graph filters (intuitive extension).** Temporal variations of the input signal can be captured by the FIR filter taking into account its temporal history. Consider the recursion

$$\boldsymbol{y}_t = \sum_{k=0}^{K} a_k \boldsymbol{L}^k \boldsymbol{x}_{t-k}, \tag{5}$$

where now the output at time $t \geq K$, i,e., $\boldsymbol{y}_t$, depends on the past $K$ realizations of the input signal, where $\boldsymbol{x}_{t-k}$ is graph-shifted with $\boldsymbol{L}^k$ (this favors a distributed implementation). Recursion (5) provides the intuition that it represents an FIR$_K$ filter in both the graph and temporal frequency domain. To see this, we calculate the joint transfer function of the filter (5). Applying first the GFT and then the $z$-transform to (5), the joint graph-temporal transfer function can be written as

$$H(z, \lambda) = \sum_{k=0}^{K} a_k \lambda^k z^{-k}. \tag{6}$$

We can now formally see that, the joint transfer function (6) implements an FIR filter of order $K$ in the graph domain, as well as, an FIR filter of the same order in the temporal domain. In a distributed computation, for computing $\boldsymbol{y}_t$ we need to access to the terms $\boldsymbol{L}\boldsymbol{x}_{t-1}, \boldsymbol{L}^2 \boldsymbol{x}_{t-2}, \ldots, \boldsymbol{L}^K \boldsymbol{x}_{t-K}$. To reduce the computation effort, we can consider that each node memorizes the terms $\boldsymbol{x}_{t-1}, \boldsymbol{L}\boldsymbol{x}_{t-2}, \ldots, \boldsymbol{L}^K \boldsymbol{x}_{t-K-1}$ while computing $\boldsymbol{y}_{t-1}$. In this way, each node can compute $\boldsymbol{L}^k \boldsymbol{x}_{t-k}$ directly from $\boldsymbol{L}^{k-1}\boldsymbol{x}_{t-k}$, which leads to the same computational effort as computing (4). From (6), we observe that the zeros of the polynomial in $\lambda$ and in $z$ are correlated to each other[2]. This affects the joint design, and thus the approximation accuracy, where a tradeoff has to be found between the filter approximations in each domain.

We illustrate this in Fig. 1, where we approximate with an FIR$_3$ filter an ideal step function in the graph frequency domain with cut-off frequency $\lambda_c = 0.5$. We can see that for a high normalized temporal frequency the filter response differs from the case of $f = 0$, for which the filter has been designed. This behavior can be addressed to the fact that the joint transfer function (6) is not a complete 2-dimensional polynomial of order $K$. Indeed, all the cross term monomials of the form $\lambda^\alpha z^{-\beta}$ with $\alpha \neq \beta$ are missing. However, even considering all these challenges, this approach can still be used to approximate some specific 2-dimensional filter masks.[3]

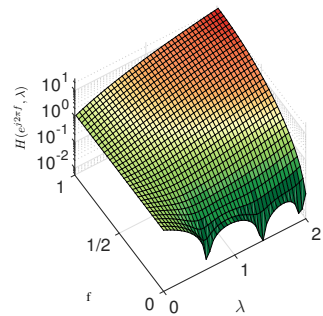

Fig. 1. The joint graph and temporal frequency response of the FIR$_3$ graph filter, designed to approximate an ideal low pass (step) response with cut-off frequency $\lambda_c = 0.5$. A normalized Laplacian has been used, $\lambda \in [0, 2]$ and the temporal normalized frequency $f \in [0, 1]$.

**General 2D FIR.** The approximation accuracy of the 2D FIR filter can be improved if we incorporate also the missing cross term monomials in (5). This can be achieved by considering all $K$ graph-shifts for every past input of the graph signal. Clearly, this approach considers more data exchanges and computational power to implement a filter of the same order in both the graph and temporal domain. More formally, consider the recursion

$$\boldsymbol{y}_t = \sum_{k=0}^{K_\text{g}} \sum_{l=0}^{K_\text{t}} a_{k,l} \boldsymbol{L}^k \boldsymbol{x}_{t-l}, \tag{7}$$

with $K_\text{g}$ and $K_\text{t}$ the memory of the filter in the graph and temporal domain, respectively. We can now calculate the joint transfer function of (7) in the same way as we did for (5). By applying the GFT and the $z$-transform we obtain

$$H(z, \lambda) = \sum_{k=0}^{K_\text{g}} \sum_{l=0}^{K_\text{t}} a_{k,l} \lambda^k z^{-l}, \tag{8}$$

which is now an FIR of order $K_\text{g}$ in the graph domain and of order $K_\text{t}$, not necessarily equal to $K_\text{g}$, in the temporal domain. From the joint frequency response (8), we can see that now we have a full polynomial in the variables $\lambda$ and $z$. Thus with this expression we can operate on all the $K_\text{g}K_\text{t}$ coefficients $a_{k,l}$, instead of the $K$ offered by (5), to approximate a given two dimensional frequency response. Further, with respect to (5) and the ARMA filter [11], recursion (7) has the potential to achieve filters of different orders in each domain. Similar to the distribution implementation of (5), the computational efforts of computing the output $\boldsymbol{y}_t$ can be reduced by allowing the nodes to keep in memory the terms $\boldsymbol{L}^k \boldsymbol{x}_{t-l}$ for $k \in [0, K_g]$ and $l \in [1, K_t]$ obtained while computing $\boldsymbol{y}_{t-1}$. Thus, for the computation of $\boldsymbol{y}_t$ only the terms $\boldsymbol{L}\boldsymbol{x}_t, \ldots, \boldsymbol{L}^{K_g}\boldsymbol{x}_t$ have to be computed. For a 2D FIR filter of order $K$ in both graph and time, this means that we need only $K$ times more computational efforts compared to (4).

One possible application of (7) is in prediction applications, where differently from the ARMA approaches [11] or the autoregressive models [12] it does not have stability issues.

---

[2]In case of ARMA filters this is observed for the poles, which affects both the filter design and stability.

[3]This holds also for the ARMA filters [11], which render both approaches suitable for graph signals that varies slowly in time ($f \approx 0$), where its graph frequency response is similar to the static case ($f = 0$).

**Causal 2D FIR.** Given the general form (7) and its particular version (5) there is room to use an intermediary approach, which can be implemented in a causal way with relaxed implementation costs. The causal 2D FIR with most degrees of freedom can be implemented as

$$\boldsymbol{y}_t = \sum_{l=0}^{K_t} \sum_{k=0}^{l} a_{k,l} \boldsymbol{L}^k \boldsymbol{x}_{t-l}. \tag{9}$$

Being a causal implementation, the terms $\boldsymbol{L}\boldsymbol{x}_t, \ldots, \boldsymbol{L}^{K_g}\boldsymbol{x}_t$ are not anymore necessary to be computed in (9), thus the distributed implementation costs are the same as (4).

**Separable 2D FIR.** In the rest of this paper, we will address a particular subclass of interest of (7) which has the property to achieve a *separable* 2-dimensional frequency response in graph and time. By setting $a_{k,l} = b_k c_l$ we can express (7) as

$$\boldsymbol{y}_t = \left(\sum_{k=0}^{K_g} b_k \boldsymbol{L}^k\right) \left(\sum_{l=0}^{K_t} c_l \boldsymbol{x}_{t-l}\right), \tag{10}$$

where $b_k$ are the filter coefficients relative to the graph part and $c_l$ to time. Similar to the derivation of (7), the transfer function of (10) can then be written as $H(z, \lambda) = H_t(z)H_g(\lambda)$ where $H_t(z) = \sum_{l=0}^{K_t} c_l z^{-l}$ and $H_g(\lambda) = \sum_{k=0}^{K_g} b_k \lambda^k$. This separable approach offers us the freedom to handle the filter specifications independently in the graph and temporal domain. We can also see (10) as first computing locally at each node the temporal filtering, i.e., $\tilde{\boldsymbol{x}}_t = \sum_{l=0}^{K_t} c_l \boldsymbol{x}_{t-l}$ and then performing the overall output $\boldsymbol{y}_t$ by filtering $\tilde{\boldsymbol{x}}_t$ on the graph, i.e., $\boldsymbol{y}_t = \sum_{k=0}^{K_g} b_k \boldsymbol{L}^k \tilde{\boldsymbol{x}}_t$. In contrast to the latter, (10) allows for an online processing of the time-varying graph signal. In the sense that, when $\boldsymbol{x}_{t+1}$ becomes available, (10) requires processing only this signal among the graph and not calculating locally $\tilde{\boldsymbol{x}}_{t+1}$ and then perform a graph filter. This means that we can track the time variations of the graph signal. However, notice that the separable filters have only $K_g + K_t$ degrees of freedom instead of $K_g K_t$ of the general case (7). Thus it can address a limited class of 2-dimensional frequency responses, but a very practical class.

**Filter design problem.** Consider a given 2-dimensional frequency mask $H^*(e^{j\omega}, \lambda)$ and we are interested in finding the filter coefficients in order to approximate it.

The design problem, for the non-separable cases, can be achieved by a 2-dimensional polynomial fitting of $H(z, \lambda)$ in the form (6) or (8) (for $z = e^{j\omega}$) to $H^*(e^{j\omega}, \lambda)$.[4] However, in case of the prediction task, where $\boldsymbol{y}_t = \boldsymbol{x}_{t+1}$, we can use some training data and design the filter coefficients directly in the graph domain by solving

$$\min_{a_{k,l}} \left\|\boldsymbol{x}_{t+1} - \sum_{l=0}^{K_t}\sum_{k=0}^{l} a_{k,l} \boldsymbol{L}^k \boldsymbol{x}_{t-l}\right\|_2^2, \tag{11}$$

for the causal model (9), or for the more general case we can substitute the causal FIR in (11) with (7).

For the case that the desired frequency response is separable, i.e., $H^*(e^{j\omega}, \lambda) = H_t^*(e^{j\omega}) H_g^*(\lambda)$ we have the benefit to separate the filter design as well. Thus, we can use any desired method to find the coefficients $b_k$ that approximate $H_g^*(\lambda)$, as well as any of the well-known techniques to find the coefficients $c_l$ for approximating $H_t^*(e^{j\omega})$. The latter renders the separable approach very practical, since we can give our specifications independently in the graph and temporal domain.

## IV. STOCHASTIC ANALYSIS

We now analyze the FIR filter behavior when the graph signal has a stochastic nature over time. This may happen, for instance, when the signal on the graph is corrupted by noise. Similar to the ARMA$_1$ graph filter [14], we characterize statistically the 2-dimensional FIR when the graph signal has a temporally non-stationary mean but a temporally stationary covariance. We consider the following signal model.

*Random signal model.* The graph signal $\boldsymbol{x}_t$ at time instant $t$ is a realization of a random process with time-varying first order moment $\bar{\boldsymbol{x}}_t$ and time-invariant covariance matrix $\boldsymbol{\Sigma_x}$. Further, the signal $\boldsymbol{x}_t$ is independent over time.

The above random signal model considers that the graph signal might be correlated among the nodes for a fixed time instant $t$, but has independent realizations with different means over time. It can be interpreted, for instance, as a desired time-varying signal $\bar{\boldsymbol{x}}_t$ embedded in noise, more specifically in the form $\boldsymbol{x}_t = \bar{\boldsymbol{x}}_t + \boldsymbol{n}_t$, with $\boldsymbol{n}_t$ being time-independent zero-mean noise with covariance matrix $\boldsymbol{\Sigma_x}$. With this in place, the following can be claimed.[5]

*Proposition 1:* Consider a separable 2D FIR filter of orders $K_g$ and $K_t$ in the graph and temporal domain, respectively, and consider a graph signal that follows the proposed random signal model. Then, the expected value $\bar{\boldsymbol{y}}_t$ and the covariance matrix $\boldsymbol{\Sigma_y}$ of the output signal are given by

$$\bar{\boldsymbol{y}}_t = \sum_{k=0}^{K_g}\sum_{l=0}^{K_t} b_k c_l \boldsymbol{L}^k \bar{\boldsymbol{x}}_{t-l} \tag{12a}$$

and

$$\boldsymbol{\Sigma_y} = \|\boldsymbol{c}\|_2^2 \left(\sum_{k=0}^{K_g} b_k \boldsymbol{L}^k\right) \boldsymbol{\Sigma_x} \left(\sum_{m=0}^{K_g} b_m (\boldsymbol{L}^m)^\top\right), \tag{12b}$$

where $\boldsymbol{c} = [c_1, c_2, ..., c_{K_t}]^\top$.

*Proof:* (Sketch) The results can be proven by applying the definitions of the expected value and the covariance matrix to the separable version of (7). ∎

Proposition 1 extends the analysis of 2-dimensional FIR filters to a stochastic environment. It says that, in the mean, the FIR filter behaves as the same 2-dimensional filter operating on a deterministic time-varying signal, being the time-varying mean of the input graph process. Further, it gives us a closed-form formula to calculate the covariance matrix of the output signal in order to see how far from the mean a given realization can be. We notice from (12b) that the variance of the output signal, at each node, depends on the squared norm of the temporal filter taps. This gives us a handle on the variance of the output signal by tuning these coefficients.

*Corollary 1:* Under the same conditions of Proposition 1 and additionally given that the covariance matrix of the input signal is diagonalizable with the graph Laplacian, i.e., $\boldsymbol{\Sigma_x} = \boldsymbol{\Phi} \mathbf{P} \boldsymbol{\Phi}^\top$, the covariance matrix of the 2D FIR output is

$$\boldsymbol{\Sigma_y} = \|\boldsymbol{c}\|^2 \boldsymbol{\Phi} \text{diag}(\hat{\boldsymbol{b}}) \mathbf{P} \text{diag}(\hat{\boldsymbol{b}}) \boldsymbol{\Phi}^\top, \tag{13}$$

---

[4]This aspect will be covered in more detail in future research.

[5]We derive the statistics for the separable case of interest, but the same can be derived also for (7) or (9).

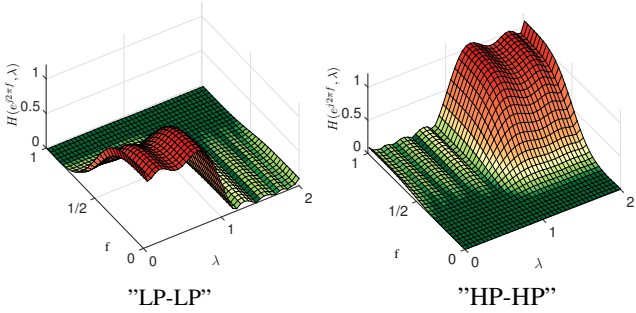

"LP-LP"  "HP-HP"

Fig. 2. Different 2-dimensional FIR approximations. From left to right, we have a low-pass (LP) filter in both graph and temporal frequency domain and a high-pass (HP) filter in both domains. A normalized Laplacian has been used and also the temporal frequencies are normalized ($\times \pi$ rad/sample).

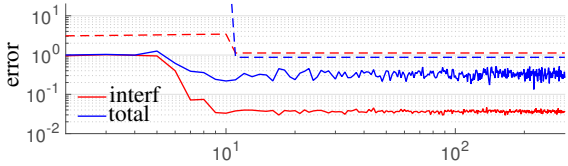

Fig. 3. Errors as a function of time for the 2-dimensional FIR graph-temporal filter (solid line) and for the classical FIR graph filter (dashed line).

where $\hat{b}$ is a vector with $n$th entry $\hat{b}_n = \sum_{k=0}^{K_g} b_k \lambda_n^k$. Corollary 1 is a straightforward extension of (12b) when $\Sigma_x$ shares the same eigenvectors of the graph Laplacian. In case the random graph signal has also a constant mean among the graph, i.e., $\bar{x}_t = \bar{x}\mathbf{1}$, we can say that (13) gives insights how the 2D FIR filter operates on the so-called power spectral density [15]–[17] ($P$) of the time-varying graph signal. In this case $\Sigma_y$ contains information on both the graph and temporal power spectral density of the output signal.

## V. NUMERICAL EVALUATION

To illustrate our conclusions, we first show that (10) can approximate different separable filters with given specifications. Then, we use the 2-dimensional FIR filter to denoise a time-varying signal which is also affected by interference. Finally, we simulate a scenario where the graph signal is a stochastic process with a time-varying mean. For the filter design, we use the polynomial approximation [9] for the graph domain and the windowing method for the time domain [18]. The results are carried over a graph of 100 nodes randomly placed in a squared area, with two nodes being neighbors if they are physically closer than 15% of the maximum distance in the area and the FIR filter of orders are $K_g = K_t = 10$.

**Filter approximation.** With reference to Fig. 2, we can see that the proposed approach can approximate different desired 2-dimensional separable frequency responses. For this particular case, the cut-off frequencies in both domains are chosen as the half of the respective bands.

**Denoising and interference cancellation.** Consider a graph signal of the form $x_t = s_t + i_t + n_t$, where

$$\langle s_t, \phi_n \rangle = \begin{cases} e^{j\pi t/4} & \text{if } \lambda_n < 0.5 \\ 0 & \text{otherwise,} \end{cases} \quad (14)$$

is the signal of interest, $\langle i_t, \phi_n \rangle = e^{j3\pi t/4}$, $\forall \lambda_n$, is the interfering signal and $n_t$ is a zero mean additive Gaussian noise with $\Sigma_{xx} = \sigma^2 I$ and $\sigma^2 = 0.1$. Our goal is to recover the graph signal of interest $s_t$ using a 2-dimensional FIR graph filtering approach. In this way, we aim to use the FIR filter to cancel the out of band noise in the graph and time domain and also suppress the interferer in the temporal domain.

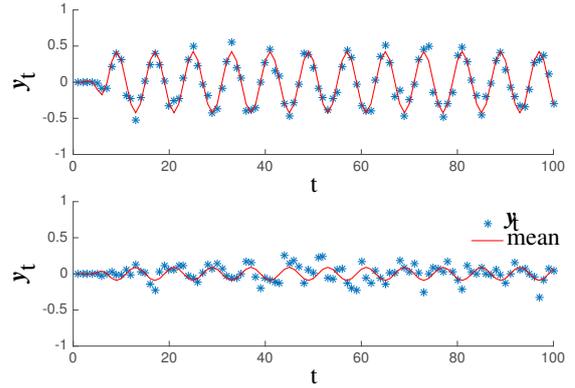

Fig. 4. The output signal of the filter and its expected value as a function of time for two nodes, node 5 (top) and node 100 (bottom).

To measure the performance of our solution, we define the following two errors

$$e_t^{(\text{interf})} = \frac{\|\hat{y}_t - \hat{y}_t^*\|}{\|\hat{y}_t^*\|}, \quad e_t^{(\text{total})} = \frac{\|\hat{y}_t - \hat{s}_t\|}{\|\hat{s}_t\|} \quad (15)$$

where $\hat{y}_t$ and $\hat{y}_t^*$ are the graph Fourier transforms of the filter outputs at time $t$ when the input signals are $x_t$ and $s_t + n_t$, respectively. The first error, $e_t^{(\text{interf})}$, is a measure on how good we attenuate the interfering signal. Indeed, it tells us how well we approximate the filter output $\hat{y}_t^*$ when there is no interference. Meanwhile, the second error, $e_t^{(\text{total})}$ tells us how good we suppress the interference and the noise. For our simulations, the coefficients $b_k$ are designed to approximate an ideal low pass step function with $\lambda_c = 0.5$, meanwhile the coefficients $c_l$ are found by approximating a low-pass temporal filter with cut-off frequency $\omega_c = \pi/2$. In Fig. 3, we can see that after some initial assessment time of the filter, both errors reduce. Specifically $e_t^{(\text{interf})}$ is reduced by an order of two[6]. This shows the robustness of the 2-dimensional FIR filter (7) to interference, where the interfering signal is attenuated by the filter. An FIR graph filter that considers the input signal only once, in the beginning of the filtering, produces errors which are much higher due to their impossibility to operate on the temporal frequencies.

**Stochastic analysis.** With the same setup as in the previous paragraph, we plot in Fig. 4 the output signal of the filter and its analytical expected value (12a) as a function of time for two representative nodes. We can notice that the output signal oscillates with the same frequency of the desired input signal around its expected value. Further, due to the fact that we attenuate the noise out of the band of interest we have also a reduction of the signal variability around its mean. The theoretical/empirical variance for node 5 is 0.0074/0.0069 and for node 100 is 0.0140/0.0149. These results show also that the variance of the output signal is reduced by more than one order w.r.t. the variance of the input signal. The empirical variance is calculated over 1000 samples.

---

[6]To further reduce the error the filter can also be designed as a band-pass around the oscillating frequency of $s_t$.